\documentclass[%
 reprint,
 amsmath,amssymb,
 aps,
pra,
]{revtex4-2}
\usepackage{url}
\usepackage{graphicx}
\usepackage{dcolumn}
\usepackage{bm}
\usepackage[colorlinks=true, urlcolor=blue, linkcolor=blue, citecolor=blue]{hyperref}

\begin{document}

\preprint{APS/123-QED}

\title{Tunneling Time for Walking Droplets on an Oscillating Liquid Surface}

\author{Chuan-Yu Hung$^{1,2}$, Ting-Heng Hsieh$^{1}$, and Tzay-Ming Hong$^{1}$ }
\thanks{ming@phys.nthu.edu.tw}
\affiliation{$^1$Department of Physics, National Tsing Hua University, Hsinchu, Taiwan 30013, Republic of China\\$^2$Department of Physics, National Taiwan University, Taipei 10617, Taiwan}\
\date{\today}

\begin{abstract}
In recent years, Couder and collaborators have initiated a series of studies on walking droplets. Experimentally, they found that at frequencies and amplitudes close to the onset of Faraday waves, droplets on the surface of silicone oil can survive and walk at a roughly constant speed due to resonance. Droplets excite local ripples from the Faraday instability when they bounce from the liquid surface. This tightly coupled particle-wave entity, although a complex yet entirely classical system, exhibits many phenomena that are strikingly similar to those of quantum systems, such as slit interference and diffraction, tunneling probability, and Anderson localization.
In this Letter, we focus on the tunneling time of droplets. Specifically, we explore (1) how it changes with the width of an acrylic barrier, which gives rise to the potential barrier when the depth of the silicone oil is reduced to prevent the generation of ripples that can feed energy back to the droplet, and (2) the distribution of tunneling times at the same barrier width. Both results turn out to be similar to the numerical outcome of the Bohmian mechanics, which strengthens the analogy to a quantum system. Furthermore, we successfully derive analytic expressions for these properties by revising the multiple scattering theory and constructing a ``skipping stone" model. 
Provided that the resemblance in tunneling behavior of walking droplets to Bohmian particles is not coincidental, we discuss the lessons for the Copenhagen interpretation of quantum mechanics that so far fails to explain both characteristics adequately.
\end{abstract}

showkeys class option if keyword
\maketitle


\section{Introduction}

It was first reported \cite{nature_walking_droplet} in 2005 that small silicone droplets can bounce indefinitely and walk at a roughly constant speed on a bowl of the same liquid due to
oscillations near the first Faraday instability \cite{Faraday1837}. Combined with the ripples it arouses at each bounce, the droplet/wave travels as a composite entity that temptingly reminds us of the wave-particle duality, an essential concept in quantum mechanics.  
Therefore, it comes as no surprise that many quantum properties have been revealed since then, such as the probability distribution of walking droplets \cite{circular_corral,square_corral}, diffraction and interference \cite{slit,slit_wrong,slit_structure}, tunneling probability \cite{expdecay}, and Anderson localization \cite{Abraham}.
After Portiere, Boudaoud, and Couder gave a detailed reason for the behavior of walking droplets \cite{particle_wave_interface},
 Moláček and Bush \cite{bush_hyd_pw} improved the droplet model to align with experimental results better in 2013.  

The analogy between walking droplets and quantum particles is not without challenges. 
The first objection was raised by Andersen {\it et al.} \cite{slit_wrong} in 2015 for the double-slit experiment by Couder and Fort \cite{slit}. They claimed that the single-particle statistics would be altered if the droplet source was moved and directed towards the midpoint of one of the slits or if a subset of the trajectories with particular impact parameters was singled out.
Just as Batelaan {\it et al.} \cite{ele_slits} want to know how the slit interacts with electrons in electron double-slits experiment, Ellegaard and Levinsen \cite{slit_structure} in 2024 aim to discover how the structure of slits affects waves and particles, further uncovering the secrets of the slit experiment involving droplets. In the meantime, the resemblance to the probability distribution in a quantum corral \cite{circular_corral} was also called into question.  After reproducing and validating the results of Ref.\cite{circular_corral}, Cheng \cite{square_corral} in 2020 replaced the rhombus container with a square one and found that the resulting probability resembled the pattern of nonlinear standing waves characteristic of the Faraday instability the system borders, instead of a checkerboard expected for an infinite square-well quantum corral.

After Eddi, Fort, Moisy, and Couder \cite{expdecay} measured the tunneling probability of a walking droplet and found its dependence on the barrier width $L$ to resemble the quantum tunneling or $\exp(-\alpha L)$ as derived via the WKB approximation \cite{shankar}, it struck us as a natural task to next record the time it takes for the particle to traverse the barrier \cite{tunneling_time_importment}. 
Set aside the stance on whether a link to the quantum system is meaningful, the tunneling time $\tau$ is an interesting property by itself. Riddled with defects and debates, the definition of $\tau$ has gone through many versions, e.g., Buttiker-Landauer time \cite{image_v_q}, phase time \cite{phase_time_spport, phase_time_1, Hartman,Hartman_Effect_dis}, and Larmor time \cite{ larmor_first,larmor_first_1,larmor_t}, and inspired many delicate and ingenious  
experiments \cite{larmor_first_exp,faster_than_light_prl,fast_then_light,steinberg,faster_than_light_more,ionization_tunneling,ele_tun_time}. 

Boosted by the consistency between experiments and theory \cite{phase_time_spport}, the phase time \cite{tunneling_mod_review}, also called Wigner–Smith time delay, is arguably the most popular definition of tunneling time in recent years. Similar concepts and discussions have been around before being popularized by Susskind and Glogower \cite{phase_time_1}. It is determined by differentiating the phase of the wavefunction at $x=L$ with respect to the energy $E$ where $0$ and $L$ mark the start and end positions of the barrier. Easy to confirm that this definition is consistent with our expectation for $V_0=0$, i.e., $L/\sqrt{2E/m}$ since $d(kL)/dE=L\cdot d\sqrt{2E/m}/dE$. 
Theoretically, the phase time is also not hard to calculate by monitoring the time duration for the peak of the transition Gaussian wave packet with mean value incident energy $E$ to enter at 0 and reappear at $L$ using the method of stationary phase \cite{group_phase}.
As calculated in Supplemental Material (SM \cite{sm}), $\tau$ becomes independent of $L$ for a wide barrier. In other words, the wavefunction has to possess a preknowledge of $L$ to finish the journey within the same time. This property was found in 1962 by Hartman \cite{Hartman}. In addition to this clear violation of causality \cite{Hartman_Effect_dis},
another problem arises, i.e., sooner or later the speed of travel will have to exceed the vacuum speed of light when $L$ gets large. 

Surely, this violation of special relativity will not happen to our system since the tunneling of walking droplets can be monitored with the naked eye every step of the way. If we want to draw an analogy to the quantum system, the definite trajectory of walking droplets will make it more suitable to compare with the prediction on tunnel time by Bohmian mechanics \cite{bohm,entropy,wiki} rather than the Copenhagen school because the latter emphasizes the wave nature, i.e., the probability distribution. It is worth noting that
Bohmian mechanics has enjoyed a renaissance in recent years as an outcome of more and more indications that the canonical teaching of quantum physics is incomplete \cite{qt_james,co_ambi} and may benefit from the insights provided by Bohm's theory \cite{ana,arrive_time,nature_bohm}.

\section{Theoretical models}
\subsection{Revised Multiple Scattering Theory}

Following the multiple scattering theory \cite{multi}, we consider quantum tunneling as the superposition of multiple pathways with different weightings. These pathways represent various ways the particle moves back and forth within the barrier. Unlike the path integral formalism \cite{feynman}, multiple scattering theory is a simplified representation that only considers the path where the walking droplet touches the edge of the barrier. When it happens, the droplet can choose between passing through or bouncing back from the edge. The ratio between them is not affected by other pathways.

After introducing the general concept of the multiple scattering theory, we add some features of our own. Let \(\hat{T}\) be the time operator, and use \(|\phi_{n}\rangle\) to denote the eigenpath that bounces  \(n\) times and obeys \(\hat{T}|\phi_{n}\rangle = \tau_{n}|\phi_{n}\rangle\) where $\tau_n$ corresponds to the time the particle spends in the barrier. The projection of the complete state $|\phi\rangle$ on \(|\phi_{n}\rangle\) can be calculated \cite{sm} as:
\begin{equation}
\langle\phi_{n}|\phi\rangle =\frac{2ik}{ik-K} \frac{2iK}{iK+k} e^{-ikL-KL} \left(\frac{-k+iK}{iK+k}e^{-KL}\right)^{2n}
\label{project}
\end{equation}
where $K = \sqrt{2m(V_0-E)}$ and $k = \sqrt{2mE}$. The probability for the particle to bounce back and forth for $n$ times can then be calculated as:
\begin{equation}
\frac{\langle\phi|\phi_{n}\rangle\langle\phi_{n}|\phi\rangle}{\sum_{m=0}^{\infty}\langle\phi|\phi_{m}\rangle\langle\phi_{m}|\phi\rangle} = e^{-4nKL}(1-e^{-4KL})
\label{porb_time}
\end{equation}
Note that Eq. \eqref{porb_time} enables us to make two predictions. First, a new definition of tunneling time as the average time for the particle to cross the barrier: 
\begin{equation}
\tau\equiv \sum_{n=0}^{\infty} \tau_{n} e^{-4nKL} (1 - e^{-4KL})
\label{sumtaun}
\end{equation}
where the travel distance $\ell_n=(2n+1)L$ for $\tau_n$. Second, the probability of $\tau_n$ decays exponentially with $n$.

\subsection{Skipping-Stone Model for Tunneling Time}

To predict how $\tau$ varies with $L$ and the speed $v_0$ of the droplet, we have yet to understand their effects on each $\tau_n$. Our inspiration is to analogize the tunneling droplet with the childhood activity of stone skipping \cite{skipstone} - throwing a flat stone across the water in such a way that it bounces off the surface. 
The fact that the droplet can survive for as long as the power of the shaking is turned on is due to the resonance. The time $T$ between successive bounces for walking droplets is known to be exactly twice the period $2\pi/\omega$ of the shaker. For a stationary droplet, John Bush \cite{bush_hyd_pw} showed that $T$ becomes equal to $2\pi/\omega$. Now let's analyze the horizontal motion of the droplet in a minimal model.
Assuming that the walking droplet loses $\alpha$ times of its kinetic energy at each bounce: $v^2_n -v^2_{n-1}=-\alpha v^2_{n-1}$. Approximating the discrete $n$ by the continuous time $t$, it becomes $dv/dt=-\gamma v$ with the solution
\begin{equation}\label{eq:v_decay}
v = v_{0}e^{-\gamma t}
\end{equation}
where $\gamma =\alpha/(2T)$ and $v_0$ denotes the initial velocity when the droplet arrives at $x=0$. 
Integrating both sides by $t$ gives
\begin{equation}
x=\frac{v_0}{\gamma}\big(1-e^{-\gamma t}\big).
\label{eq:x_t}
\end{equation}
Now let's go back to the tunneling particle that may bounce back and forth in the barrier for $n$ times where $n=0, 1\dots$ is discrete. It is straightforward to obtain
\begin{equation}
\tau_n = -\frac{1}{\gamma}\ln(1-\frac{\ell_n\gamma }{ v_{0}}).
\label{taun}
\end{equation}
Plugging Eq. \eqref{taun} in Eq. \eqref{sumtaun} will give the tunneling time.

\section{Tunneling Experiment}
 
The tunneling experiment is very sensitive to environmental conditions. An air conditioner was used to maintain the temperature at 25$^{\rm o}$C, but it led to airflow. So we shielded the whole setup with cardboard. Additionally, the relative humidity is kept between 40-60\%.
To simulate the energy barrier, we place an acrylic plate with a thickness of 4.00 mm across the short diagonal of a rhombus-shaped container whose width and length are 12.3 cm and 4.47 cm, respectively. Denoted by $L$, the width can be chosen among 3.20, 4.06, 4.50, 5.00, 5.46, 6.00, and 6.98 mm. We purposefully adjusted the amount of silicone oil with a viscosity of 50 cST in the container so that the region above the shallow water is far from the Faraday instability. For the record, the depth of oil in the deep and shallow areas is 6.52 and 2.52 mm. 
Mimicking the parameter ranges adopted by Ref.\cite{expdecay}, the container is placed on a shaker with the acceleration and frequency of 4.22 $g$ and 50 Hz where $g$ = 9.8 $m/s^2$. 
 
 If the oil depth is far from the conditions for the Faraday waves, the droplet will not be able to move or walk. 
This is why we expect the shallow region to function as an energy barrier that prevents the walking droplet from crossing. 
However, as reported by Ref.\cite{expdecay}, there is still a nonzero probability for the droplet to tunnel to the other side of deep oil, resembling the tunnel effect of a quantum particle. The purpose of adopting a rhombus-shaped container is to ensure the direction in which the droplets hit the barrier is mostly consistent and, as a result, they all approach the margin of the barrier at approximately a vertical angle.

With the aid of a camera (Canon EOS 850D) and the Python package OpenCV, the trajectory of walking droplets can be recorded and analyzed. 
By extracting data from the video consisting of the positions of droplets at each point in time, we can deduce the speed. Two numbers of interest here are (1) the velocity as the incident droplet touches the front edge of the barrier and (2) the time it stays in the barrier. Since the velocity of the incident droplet can vary, we classify the behavior for different incident velocities separately. As in Ref.\cite{expdecay}, the tunneling probability is defined as the ratio of crossing frequency and the total number of trials.

\section{Experimental and numerical Results}

\begin{figure}
  \includegraphics[width=0.5\textwidth]{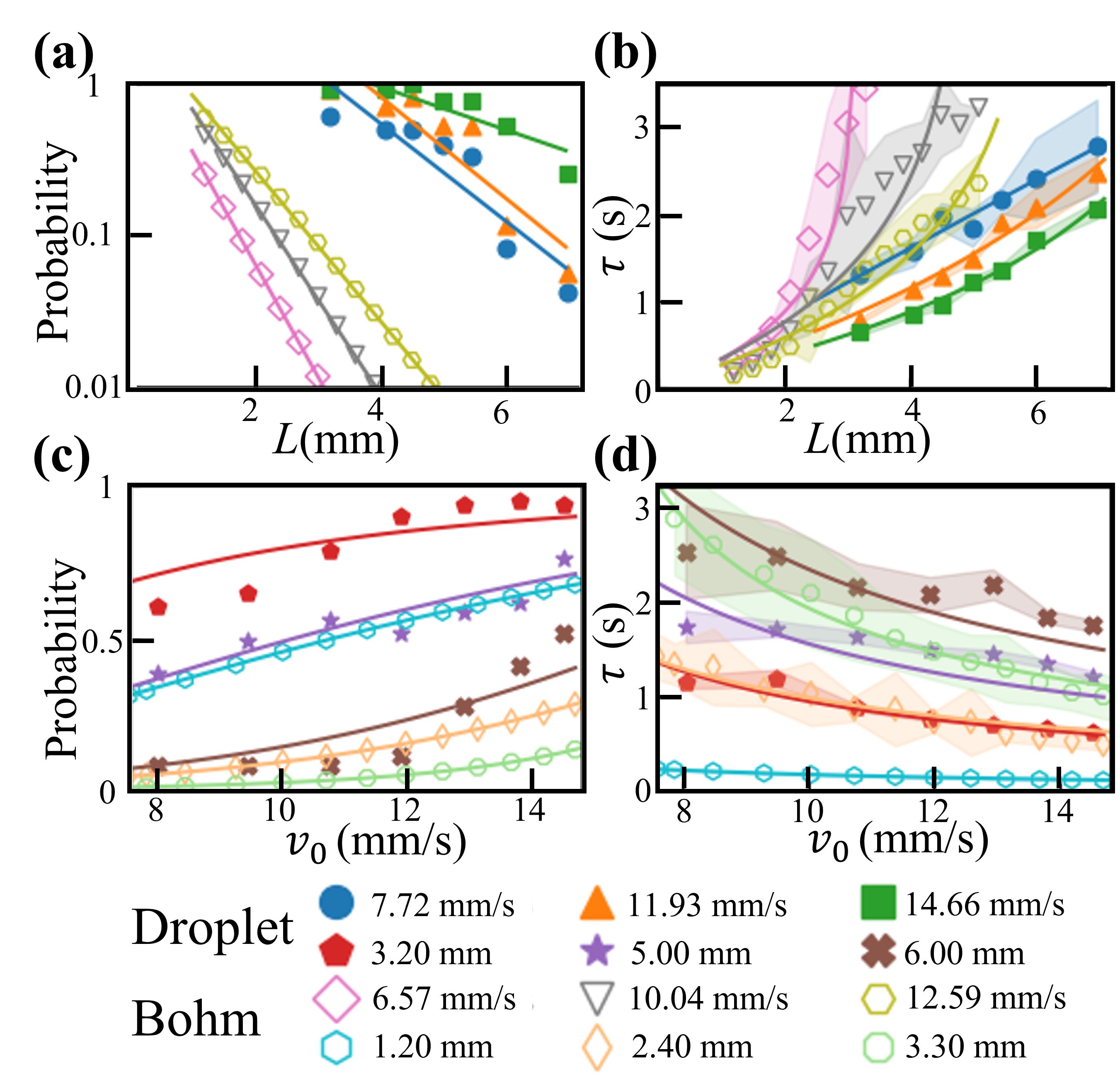} 
  \caption{(a, b) The probability and tunneling time for a droplet/particle with fixed velocity \(v_0\) are plotted as a function of \(L\). (c, d) The same measurements are done for a fixed \(L\) while changing \(v_0\). Solid lines in (a), (c), and (b, d) are theoretical predictions via the WKB approximation, full tunnel probability, and Eq. \eqref{sumtaun}. The light-colored band represents the 95\% confidence interval of the data \cite{confi}.} 
  \label{fig:combine4to1} 
\end{figure}

As mentioned earlier, we plan to compare the tunneling time of walking droplets with that of Bohmian particles. The potential barrier for the Bohmian mechanics system is $2.88 \times 10^{-5} \, \text{J}$, and we set $\hbar = 1$ and the particle mass to $0.2 \, \text{kg}$. 
To make sure that our setup is trustworthy, we first reproduced the tunneling probability by Couder {\it et al.} \cite{expdecay} in Fig. \ref{fig:combine4to1} (a, c), before recording the time spent by the walking droplets in crossing in Fig. \ref{fig:combine4to1} (b, d). We then added the numerical results by Bohmian mechanics to Fig. \ref{fig:combine4to1} and tried to fit them by the theoretical predictions in Sec. II. The fact that the same theoretical descriptions, i.e., the revised Multiple Scattering Theory and Skipping-Stone model, can reasonably well explain the experimental and numerical results simultaneously strengthens the link of the classical walking droplet to the quantum system as described by Bohm. 

\begin{figure}
  \includegraphics[width=0.5\textwidth]{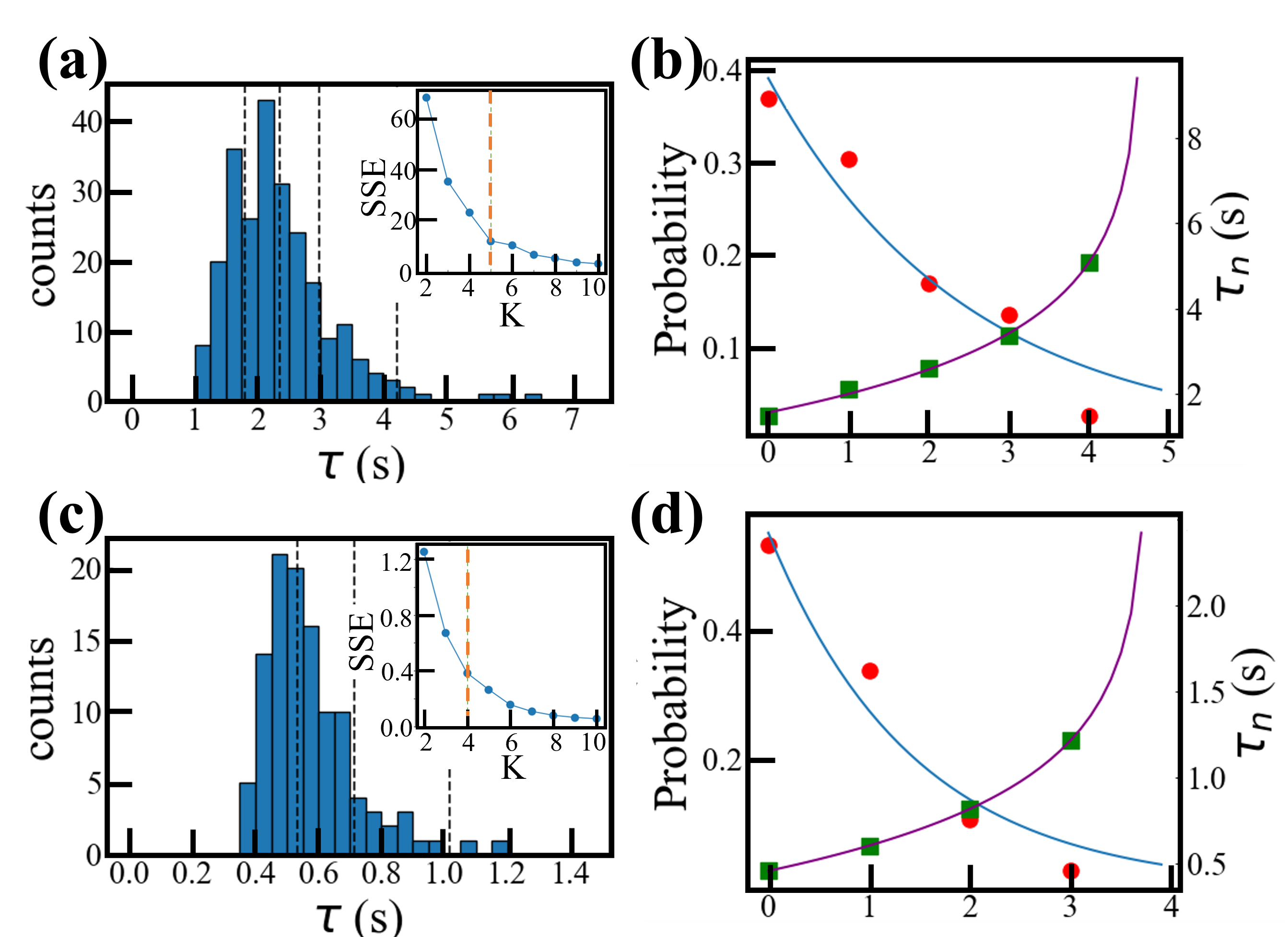} 
  \caption{(a) The distribution of $\tau$ for droplets with $v_0$ = 12.4 mm/s to cross the barrier with $L$ = 5 mm. The knowledge that the data are best divided into 5 clusters is via the K-means and elbow methods. Inset shows how 
 the sum of square error (SSE) in Eq. (C1) varies with K where the dashed line indicates the elbow point. (b) The probability weight and $\tau_n$ value for each $n$ are denoted by the red dots and green squares in this double-$y$ plot for the same $L$ and $v_0$ as in (a). The solid fitting lines are theoretical predictions from Eqs. \eqref{porb_time} and \eqref{taun}. Similar to (a, b), (c, d) are numerical results for a Bohmian particle with $v_0$ = 13.7 mm/s and $L$ = 2.40 mm.} 
  \label{fig:distribution} 
\end{figure}

\begin{figure}
  \includegraphics[width=0.5\textwidth]{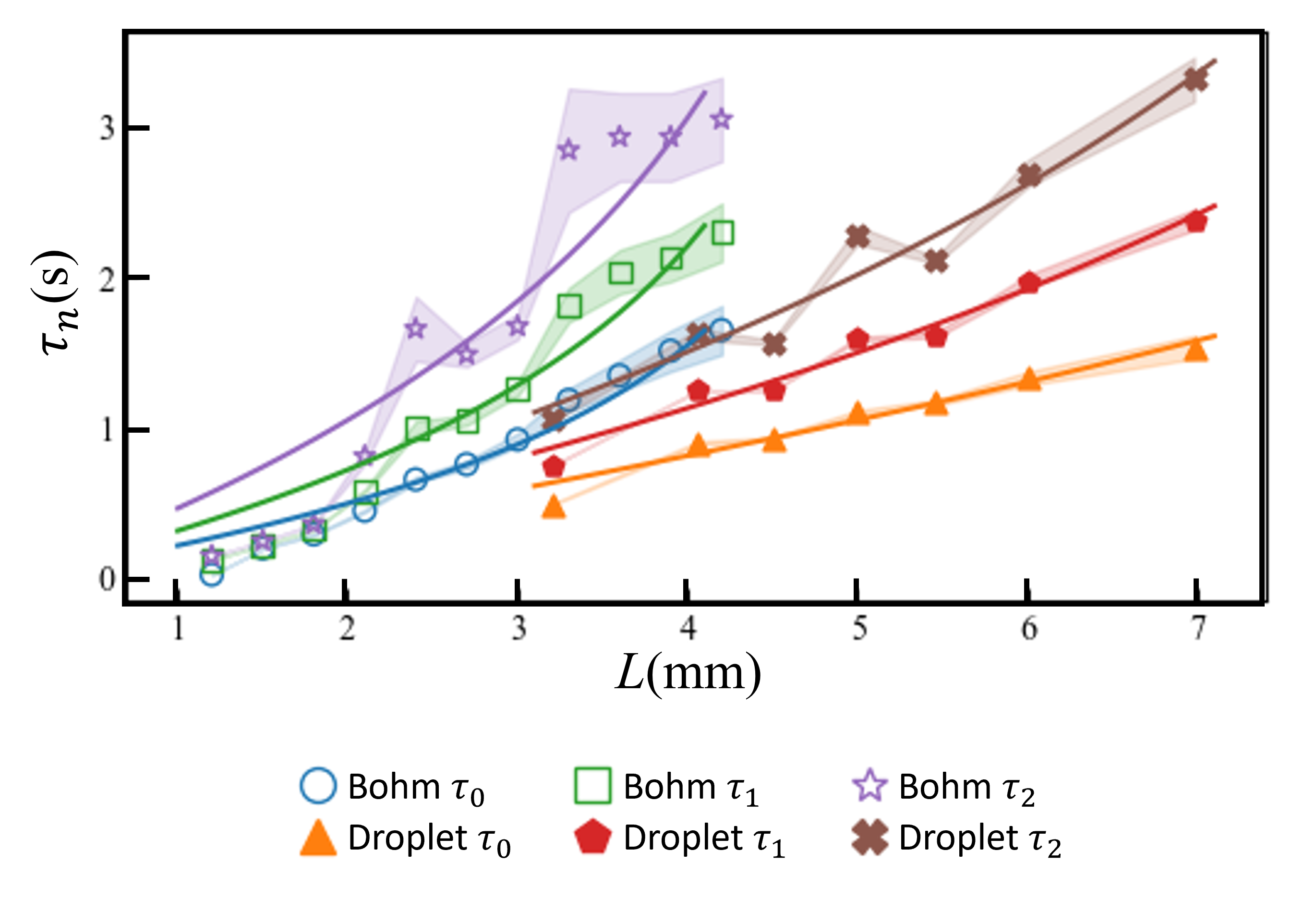} 
  \caption{The value of $\tau_n$ determined by K-means algorithm is plotted against different $L$. The $v_0$ for droplets and  Bohmian particles are 11.4 and 13.7 mm/s. The solid fitting line is the theoretical prediction in Eq. \eqref{taun}. The $\tau_3$ is not included because its probability weight is too small to allow a meaningful determination of its value.} 
  \label{fig:taunvst} 
\end{figure}

To analyze the histogram of tunneling time in Fig. \ref{fig:distribution}(a, c) objectively, we employ the statistical method of K-means \cite{kmeans} to perform the partitioning and determine the number of partitions in the distribution by using the elbow method \cite{elbowmethod}. The results, plotted in Fig. \ref{fig:distribution} (b, d), turn out to match our theoretical predictions in Eqs. \eqref{porb_time} and \eqref{taun}. 
To check the $L$-dependence of each $\tau_n$ in Eq. \eqref{taun}, we apply the same algorithm for a fixed incident velocity with different $L$ and obtain Fig. \ref{fig:taunvst}. Again, the walking droplets and Bohmian particles share the same behavior that can be captured by our toy model of skipping stones.

\begin{figure}
  \includegraphics[width=0.5\textwidth]{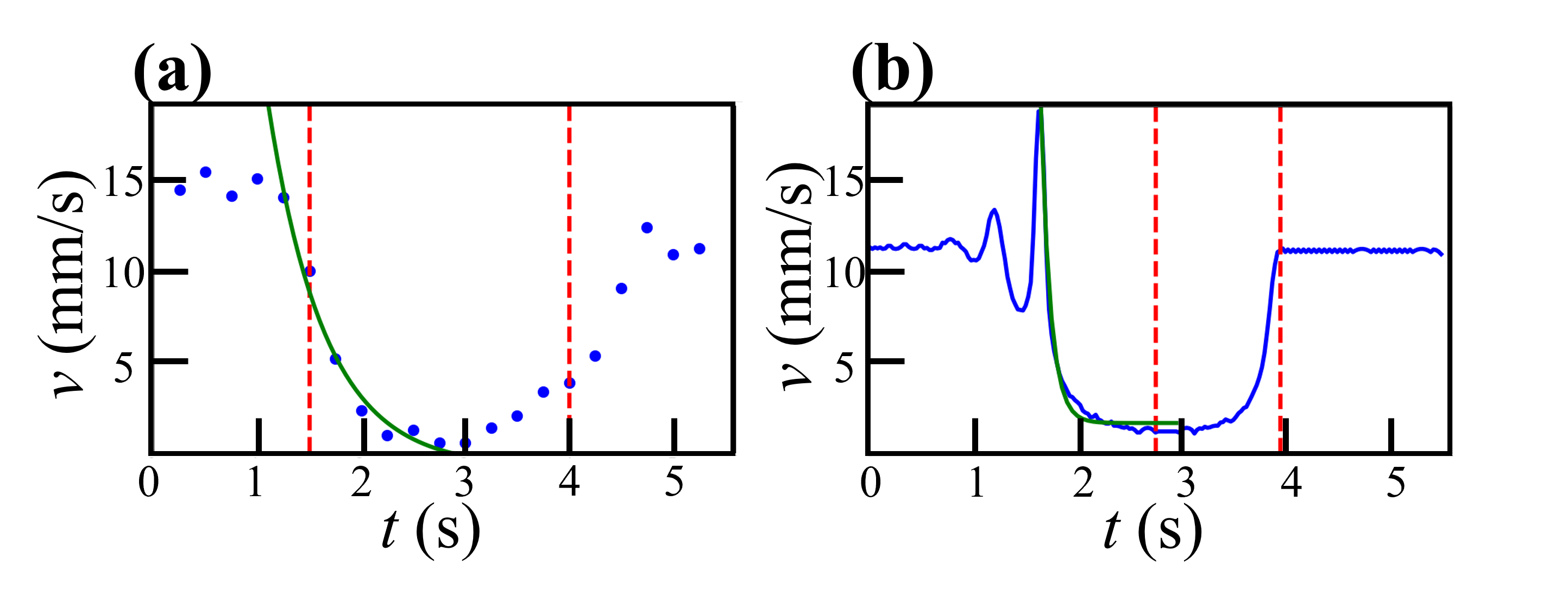} 
  \caption{(a) The change of velocity with time for droplets with $v_0=$ 14.8 mm/s tunnels through an energy barrier of $L=$ 6.00 mm. (b) Bohmian particles with $v_0=$ 10.7 cross the barrier of $L=$ 3.00 mm. The vertical dashed lines mark the time when the droplet/particle enters and exits the barrier. The green solid line denotes an exponential decay function as predicted by Eq. \eqref{eq:v_decay}.} 
  \label{fig:vvst} 
\end{figure}

Although  Eq. \eqref{taun} has successfully predicted the behavior in Figs. \ref{fig:distribution} and \ref{fig:taunvst}, there is no guarantee for the correctness of Eq. \eqref{eq:v_decay}, which describes 
how the velocity changes with time during tunneling and 
is crucial to deriving Eq. \eqref{taun}. 
This is verified in Fig. \ref{fig:vvst} (a, b) for a walking droplet and Bohmian particle.
Note that the increase in $v$ before exiting the barrier is due to the requirement that the speed return to its original speed in the deep region.  
We believe the reason why the Bohmian particle accelerates earlier is that the ripple that accompanies the walking droplet was created by its previous bounce and, therefore, always lags behind the droplet. In contrast, no such constraint on the probability wave that drives the Bohmian particle. 

\section{Conclusion and Discussions}

We set out to measure how the tunneling time $\tau$ of a walking droplet (1) varies with its initial velocity and the barrier width $L$, and (2) distributes in its magnitude, while bearing in mind to test the plausibility of making an analogy with the quantum system. The correspondence is motivated by the fact that (i) the connection has been built and based on several physical properties, (ii) the definition of tunneling time is a pending issue in quantum mechanics that has gone through various versions amid heated arguments since the advent of quantum theory, and (iii) the prevailing phase time in recent years suffers the defect of violating the speed limit of Special Relativity, and the walking droplet offers a promising alternative because its movement can be traced visually and will surely never exceed $c$.

In conjunction with the experiment, we modified the Multiple Scattering theory and devised a skipping-stone model to successfully derive analytic expressions for   
(a) the distribution of $\tau_n$,
(b) $L$-dependence of both $\tau$ and $\tau_n$,
(c) how $\tau$ shortens with increasing incident velocity $v_0$,
and (d) how the velocity changes with time.
To compare with the quantum behavior, we resorted to the Bohmian mechanics and calculated its prediction for these quantities. To our surprise, they are very similar to those of walking droplets. Provided that the resemblance is not fortuitous, the walking droplet in the shallow region allows us to visualize the tunneling behavior of a Bohmian particle. With the growing interest in Bohmian mechanics or pilot wave theory, we kept an open mind to seek insights from imagining the corpuscular nature of a tunneling quantum particle as a decelerating stone that skims through the surface of a lake.

In passing, there exists a maximum distance the stone can travel, i.e., $v_0/\gamma$ by setting $t\rightarrow \infty$ in Eq. \eqref{eq:x_t}. 
This may echo the rare observation of the walking droplet wiggling locally as if trapped in some confined region within the range of shallow oil. How a tunneling quantum particle behaves in the rare event of a thick barrier $L$ may warrant investigations in the future.

\begin{acknowledgments}
We are grateful to Chao-Lin Hsu, Hong-Yue Huang, Ya-Chang Chou, and Eric Jones for technical assistance and advice, and to the National Science and Technology Council in Taiwan for financial support under Grants No. 112-2112-M007-015 and 113-2112-M-007-008.
\end{acknowledgments}

\end{document}